\documentclass[aps,twocolumn,showpacs,superscriptaddress,preprintnumbers,floatfix,amsmath,amssymb,eufrak,table]{revtex4-1}
\pdfoutput=1
\usepackage{graphicx}
\usepackage{dcolumn}
\usepackage{bm}
\usepackage{slashed}
\usepackage{amsmath,graphicx}
\usepackage[colorlinks=true,linktocpage=true,linkcolor=blue,citecolor=blue]{hyperref}
\usepackage{float}
\usepackage{nicefrac}
\usepackage[normalem]{ulem}
\usepackage{amsmath}
\usepackage{subfigure} 
\usepackage{bbold} 
\usepackage[makeroom]{cancel}
\usepackage{stmaryrd} 

\def\bea {\begin{eqnarray}}
\def\eea {\end{eqnarray}}
\def\mn {\mu\nu}
\newcommand{\be}{\begin{equation}}
\newcommand{\ee}{\end{equation}}
\newcommand{\ba}{\begin{eqnarray}}
\newcommand{\ea}{\end{eqnarray}}
\newcommand{\nn}{\nonumber\\}
\def\be {\begin{equation}}
	\def\ee {\end{equation}}

\DeclareGraphicsExtensions{.jpg,.pdf,.eps}

\begin{document}

\title{Energy loss of a fast moving parton in Gribov-Zwanziger plasma}

\author{Manas Debnath}
\email{manas.debnath@niser.ac.in}
	\affiliation{School of Physical Sciences, National Institute of Science Education and Research, An OCC of Homi Bhabha National Institute,  Jatni, Khurda 752050, India}

\author{Ritesh Ghosh}
\email{riteshghosh1994@gmail.com}
\affiliation{College of Integrative Sciences and Arts, Arizona State University, Mesa, Arizona 85212, USA}

\author{Mohammad Yousuf Jamal}
\email{mohammad@iitgoa.ac.in}
\affiliation{School of Physical Sciences, Indian Institute of Technology Goa, Ponda 403401, Goa, India}

\author{Manu Kurian}
\email{mkurian@bnl.gov}
\affiliation{RIKEN BNL Research Center, Brookhaven National Laboratory, Upton, New York 11973, USA}

\author{Jai Prakash}
\email{jai183212001@iitgoa.ac.in}
\affiliation{School of Physical Sciences, Indian Institute of Technology Goa, Ponda 403401, Goa, India}

\begin{abstract}
{The Gribov-Zwanziger prescription applied within Yang-Mills theory is demonstrated to be an efficient method for refining the theory's infrared dynamics. We study the collisional energy loss experienced by a high-energetic test parton as it traverses through the Grivov plasma at finite temperature.  To achieve this, we employ a semi-classical approach that considers the parton's energy loss while accounting for the back-reaction induced by the polarization effects due to its motion in the medium.  The polarization tensor of the medium is estimated within a non-perturbative resummation considering the Gribov-Zwanziger approach. The modification of the gluon and ghost loops due to the presence of the Gribov parameter plays a vital role in our estimation. We observe that the non-perturbative interactions have a sizable effect on the parton energy loss. 
Further, we discuss the implications of our findings in the context of relativistic heavy-ion collisions.}\\

\end{abstract}
	\maketitle 
	\newpage


{\bf \emph{Introduction}}-{The vibrant research programs at the Relativistic Heavy Ion Collider (RHIC) and Large Hadron Collider (LHC) offer an effective way to study extremely dense and hot matter, referred to as the quark-gluon plasma (QGP), which is governed by the laws of strong interaction. These experiments yield valuable insights about the QGP's properties through the analysis of various experimental observables, such as particle spectra, anisotropic flow, etc. ~\cite{Romatschke:2007mq, Ryu:2015vwa}. High energy partons that originate from the initial hard scatterings of the nuclei serve as {\it hard probes} and play an efficient role in unraveling the properties of the QGP. They traverse through the QGP while interacting with the medium constituents and leave behind imprints of the QGP on its experimental observables \cite{vanHees:2005wb, JETSCAPE:2017eso, Das:2009vy, Cao:2018ews, Song:2020tfm, Singh:2023smw}.}

{ 
The behavior of energy loss of the moving parton is linked to the characteristics of the surrounding medium. Efforts have been made to study the complex dynamics of pre-equilibrium, unstable, expanding, magnetized, and chiral media by investigating the energy loss of partons within their respective environments~\cite{Dumitru:2007rp, Carrington:2015xca, Prakash:2021lwt,Prakash:2023wbs, Kurian:2019nna, Hauksson:2021okc, Shaikh:2021lka, Ruggieri:2022kxv, Carrington:2022bnv, Ghosh:2023ghi, 10.1088/1361-6471/ad10c9,Carignano:2021mrn}. The moving parton undergoes interactions with both hard components, arising from elastic collisions with the medium constituents carrying momenta on the order of $T$ (the temperature scale), and soft modes, which encompass the gauge fields within the medium carrying momenta around $gT$, where $g$ represents the coupling constant. The impact of the soft modes has received comparatively less attention in the literature, which may be due to the fact that the energy loss caused by the hard components is typically more significant. Nevertheless, it is essential to consider that the soft modes, represented by the gauge fields in the medium, play a non-negligible role in the interaction dynamics with the test parton. This is due to the high occupation number of these soft modes within the plasma~\cite{Carrington:2016mhd}, which leads to a non-trivial interaction frequency between the parton and the classical fields.} 

{A semi-classical approach is employed to analyze the soft contribution of energy loss experienced by an energetic parton that takes into account its interaction with the chromodynamic fields by quantifying the polarization effects of the QCD medium due to the passage of the parton. It is important to highlight that there exists a substantial disparity between the resummed perturbative results and the lattice-based estimations. For instance, a notable discrepancy has been observed in the estimation of the charm quark diffusion coefficient when comparing results from pQCD analysis with the lattice data ~\cite{Dong:2019byy}. In a very recent study~\cite{Madni:2022bea}, the authors have shown a remarkable improvement over the pQCD estimate of diffusion coefficient with the implementation of the Gribov-Zwanziger prescription~\cite{Gribov:1977wm, Zwanziger:1989mf} in the analysis. Notably, the Gribov-Zwanziger approach improves the infrared (IR) behavior of QCD, where the conventional resummed perturbative approach may not be reliable as the region is strongly coupled. This is done by addressing residual gauge transformations that persist by applying the Faddeev-Popov quantization procedure. Recent progress on the Gribov-Zwanziger approach and its application in various processes in QCD can be found in Refs.~\cite{Vandersickel:2012tz,Dudal:2008sp, Capri:2016aqq,Gotsman:2020mpg,Justo:2022vwa,Canfora:2013kma,Fukushima:2013xsa,Wu:2022nbv,Golterman:2012dx,Bandyopadhyay:2015wua,Su:2014rma,Sumit:2023hjj,Debnath:2023dhs,Bandyopadhyay:2023yjp}.}

{In this work, utilizing the Gribov-Zwanziger prescription for the first time in the analysis of the passage of a fast moving parton, we show that the energy loss is highly influenced by the non-perturbative effects and has an important role in the phenomenological implication within the context of heavy-ion collision. We construct the gluon self-energy for the Gribov plasma and determine the effective gluon propagator to quantify the induced current generated by the moving parton in the medium. Further, we formulated the energy loss of parton with the Gribov-Zwanziger approach and studied the momentum evolution and nuclear suppression factor of the test parton in Gribov plasma. }\


{
{\bf \emph{Energy loss in Gribov plasma}}-As a high-energy test parton moves through the plasma, it is subject to energy loss stemming from interactions with the color fields.  To comprehensively describe the dynamics of a test parton in the presence of these chromodynamic fields, the Wong equations offer a valuable and Lorentz covariant framework as follows~\cite{Wong:1970fu},
\begin{align}
\frac{dX^{\mu}(\tau)}{d\tau} &= V^{\mu}(\tau),\label{eq:1_1} \\
\frac{dQ^{\mu}(\tau)}{d\tau} &=g \tilde{q}^{a}(\tau)F^{\mu\nu}_{a}(X(\tau)){V}_{\nu}(\tau),\label{eq:1_2} \\
\frac{d\tilde{q}^{a}(\tau)}{d\tau} &= -gf^{abc}V_{\mu}(\tau)A^{\mu}_{b}(X(\tau))\tilde{q}_{c}(\tau),
\label{eq:wong}
\end{align} 
in which $\tau$, $X^\mu (\tau)$, $Q^\mu (\tau)$, and $V^\mu (\tau)$ correspond to the proper time, position, momentum and velocity of the test parton, respectively with a color charge $\tilde{q}_a$. Here, $F^{\mu\nu}$ denotes the chromodynamic field tensor, $A^\mu$ represents the gauge potential, $f^{abc}$ is the structure constant of the ${SU}(N_c)$ group, and $a$ serves as the color index, where $a=1,2,\dots, N_c^2-1$. These equations provide a formalism for understanding how the test parton behaves within the intricate interplay of forces and fields, shedding light on the energy loss phenomenon in this context. This energy loss is quantified by examining the work done by retarding forces acting on the parton within the medium. These forces arise from the induction of a chromo-electric field due to the parton's motion. By adopting the well-established formalism described in Refs.~\cite{Ghosh:2023ghi, Carrington:2015xca}, Wong equations, along with the linearized Yang-Mills equation, provide the parton energy loss in the Gribov plasma as,
\begin{align}
{\frac{d{ E}}{dx}}=i\frac{1}{|{\bf v}|}g^2C_F v^i v^j \int{\frac{d^3{\bf p}}{(2\pi)^3}\omega \Delta^{ij}},
\label{main}
\end{align}
with $C_F$ as the Casimir invariant of $SU(N_c)$, $\omega = {\bf p}\cdot{\bf v}$ where ${\bf v}=\frac{{\bf q}}{E_q}$ and $\Delta^{ij}$ is the IR improved gluon propagator
with the Gribov-Zwanziger approach.
The properties of Gribov plasma are captured in $\Delta^{ij}$ and can be represented by employing the Dyson-Schwinger equation as,
\bea
\Delta_{\mu\nu}^{-1}&=&\left(\Delta_{\mu\nu}^0\right)^{-1}-\Pi_{\mu\nu}.
\label{DS}
\eea
The gluon propagator with Gribov term can be described as~\cite{Su:2014rma},
\bea
\Delta_{\mu\nu}^0(P)=\left(\delta_{\mu\nu} -(1-\xi) \frac{P_\mu P_\nu}{P^2}\right)\frac{P^2}{P^4+\gamma_G^4},
\label{delta00xi}
\eea
where $P^\mu\equiv (p_0=\omega, {\bf p})$, $\xi $ denotes the gauge  parameter and $\delta_{\mn}=\text{diag}(1,1,1,1)$.  Here, $\gamma_G$ represents the Gribov parameter. Its temperature dependence can be extracted within the framework of finite-temperature Yang-Mills theory by solving the gap equation. The presence of the parameter $\gamma_G$ in the denominator has the effect of shifting the pole of the gluon to an unphysical position, specifically $P^2=\pm i \gamma_G^2$. These unphysical excitations, which emerge when incorporating the Gribov parameter, signify the effective confinement of gluons~\cite{Vandersickel:2012tz,Dudal:2008sp,Capri:2016aqq}. The gluon self-energy is characterized by two independent symmetric tensors, which can be expressed in terms of form factors $\Pi_L$ and $\Pi_T$ as,
\bea
\Pi^{\mn}= \Pi_T A^{\mn}+\Pi_L B^{\mn}.
\label{Pi}
\eea
Here, $B^{\mn}= \frac{\bar u^\mu \bar u^\nu}{\bar u^2}$ where $\bar u^\mu=G^{\mn}u_\nu$ with $G^{\mn}=\delta^{\mn}-\frac{P^\mu P^\nu}{P^2}$ and $A^{\mn}=G^{\mn}- B^{\mn}$. Employing Eq.~\eqref{delta00xi} and Eq.~\eqref{Pi} in Eq.~\eqref{DS}, the inverse effective gluon propagator can be expressed as,
\begin{align}
\Delta_{\mu\nu}^{-1}
=&\frac{P^4+\gamma_G^4}{P^2\xi}\delta_{\mu\nu}+\left(\frac{\xi-1}{\xi}\frac{P^4+\gamma_G^4}{P^2}-\Pi_T\right)A_{\mu\nu}\nonumber\\
&+\left(\frac{\xi-1}{\xi}\frac{P^4+\gamma_G^4}{P^2}-\Pi_L\right)B_{\mu\nu}.
\end{align}
The general structure of the gluon propagator can be decomposed in the tensor basis as,
\be
\Delta_{\mu\nu} = \alpha P_\mu P_\nu + \beta A_{\mu\nu} + \gamma B_{\mu\nu},
\ee
where the coefficients $\alpha, \beta,\gamma$ can be determined from the relation $\delta^\nu_\alpha=\Delta^{\mu\nu} \left(\Delta_{\mu\alpha}\right)^{-1}$. We obtain the gluon propagator in Gribov plasma as, 
\begin{align}
  \Delta_{\mu\nu} =& \frac{\xi P_\mu P_\nu}{P^4+\gamma_G^4} +\frac{P^2 A_{\mu\nu}}{P^4+\gamma_G^4-P^2\Pi_T} \nonumber\\
&+\frac{P^2B_{\mu\nu}}{P^4+\gamma_G^4-P^2\Pi_L}. \label{gluonprop} 
\end{align}
The modified dispersion relation of gluons with the Gribov term can be extracted from the pole of the propagator. It is important to emphasize that in the case $\gamma_G=0$, Eq.~\eqref{gluonprop} reduce back to the form of thermal gluon propagator. 
\begin{figure}
 \centering
\includegraphics[scale=0.55]{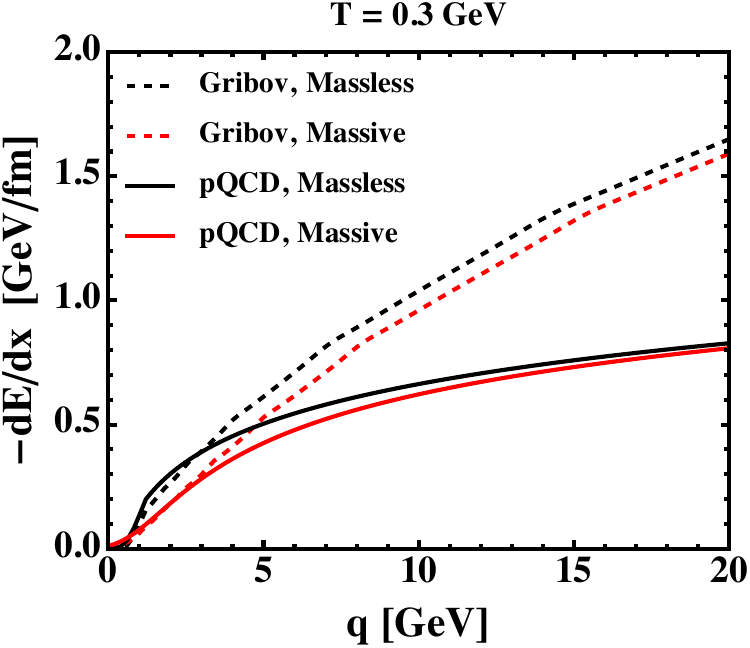}
\caption{Gribov-modified energy loss of massless and massive parton and its comparison with that from the pQCD calculations.}
\label{elp}
\end{figure}
The form factors $\Pi_L$ and $\Pi_T$ can be obtained from one loop gluon self-energy as,
\begin{align}
&\Pi_L(\omega,p)=-\frac{\omega^2-p^2}{p^2}\Pi_{00}(\omega,p),\\
&\Pi_T(\omega,p)=\frac{1}{2}[\Pi_{\mu\mu}(\omega,p)-\Pi_{L}(\omega,p)].
\end{align}
One needs to estimate the quark, gluon, and ghost loops contributions to the gluon self-energy. It is important to emphasize that the gluon and ghost loops will be affected by the Gribov-Zwanziger approach. The Gribov constraint does not affect the quark sector, as it only affects the gauge sector of the QCD. 
%
\begin{figure}[h]
 \centering
\includegraphics[scale=0.4]{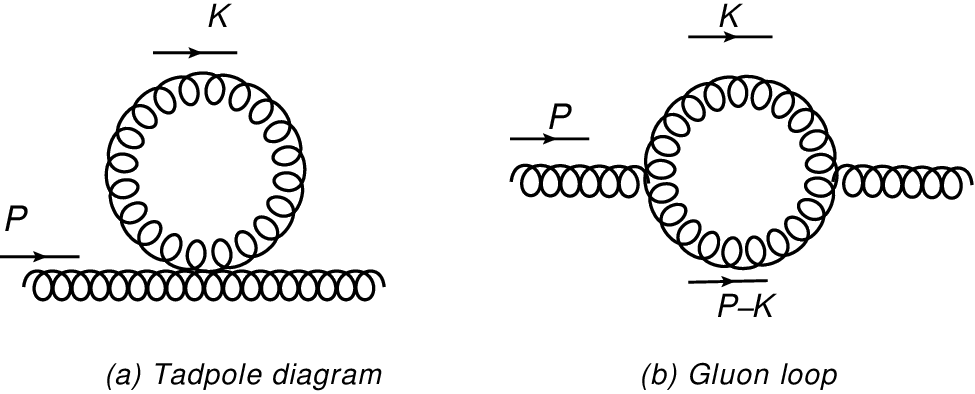}
\label{d1}
\end{figure}
We focus on the estimation of $\Pi_{\mu \mu}$ and $\Pi_{00}$ from each sector to obtain Gribov parameter dependence of $\Pi_T$ and $\Pi_L$. 

The contribution of the tadpole diagram and gluon loop to the gluon self-energy can be described as,
\begin{align}\label{gluon_tadepole1}
\Pi_{00}^{a+b}(p_0,{\bf p})&= -\frac{13}{2}g^2 C_A J_1+ 3 g^2 C_A J_{4}(p_0, {\bf p})\nonumber\\
&+ \frac{13}{2}g^2 C_A J_2(p_0, {\bf p})+6g^2 C_A J_3(p_0,{\bf p}),\\
 \Pi_{\mu \mu}^{a+b}(p_0,{\bf p})&=\frac{5}{2} g^2 C_A (\delta_{\mu \mu}-1) J_1+ \frac{1}{2}g^2 C_A \delta_{\mu \mu} J_2(p_0, {\bf p})\nonumber\\
&- 6g^2 C_A \Big(J_1-J_2(p_0, {\bf p})\Big)\label{gluon_tadepole2}.
\end{align}
It is important to note that the $\Pi_{\mu \nu}^a(P)$ is independent of external momentum. A detailed derivation of Eq.~\eqref{gluon_tadepole1} and Eq.~\eqref{gluon_tadepole2} along with the functional forms of $J_i (i=1, .., 4)$ is discussed in Section A of the supplementary material.

\begin{figure}[h]
 \centering
\includegraphics[scale=0.3]{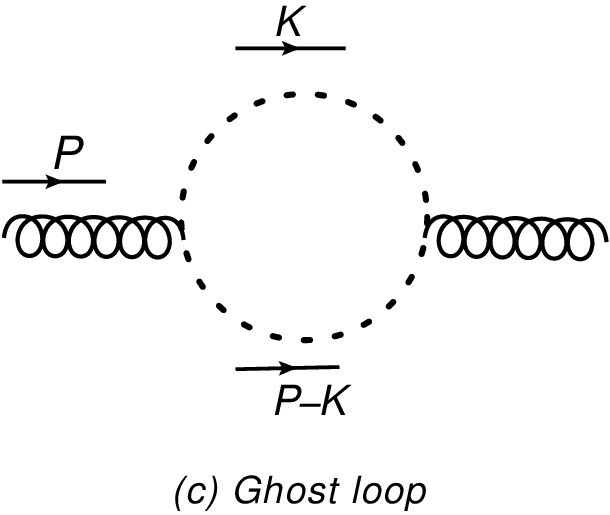}
\label{d2}
\end{figure}
In the Landau gauge, the ghost propagator in the infrared limit reads as~\cite{Vandersickel:2012tz},
\bea
D^c(K^2)=\frac{128 \pi^2 \gamma_G^2}{3g^2}\frac{1}{K^4}.
\eea
Employing the form of the ghost propagator, we can define the ghost loop contribution to $\Pi_{\mu \mu}$ and $\Pi_{00}$ as,
\begin{equation} \label{eq.2}
\Pi^c_{\mu \mu}= \left(\frac{128 \pi^2 \gamma_G^2}{3 g^2} \right)^2 I^c_{\mu \mu},  \qquad \Pi^c_{00}=\left(\frac{128 \pi^2 \gamma_G^2}{3g^2 } \right)^2 I^c_{00},
\end{equation}
where the functional form of $I^c_{\mu \mu}$ and $I^c_{00}$ take the forms as follows,
\begin{align}\label{main1}
 I^c_{\mu \mu}&=  -\frac{1}{4\pi^2} \frac{7 \zeta (3)}{8 \pi^2 T^2}\bigg[\frac{1}{4}+
 \bigg\{\frac{p_0^2}{p_0^2-p^2}-\frac{p_0}{2p}\log \left(\frac{p_0+p}{p_0-p}\right)\bigg\}\bigg],\\
I^c_{00}&=  - \frac{1}{8 \pi^2} \frac{7 \zeta(3)}{8 \pi^2 T^2}+\frac{1}{(2\pi)^2}\frac{p_0^2}{(p^2- p_0^2)^2}\nonumber\\
&- \frac{2}{(2\pi)^2} \frac{7 \zeta(3)}{8\pi^2 T^2} \bigg\{\frac{p_0^2}{p_0^2-p^2}- \frac{p_0}{2p}\log \left(\frac{p_0+p}{p_0-p}\right)\bigg\}\label{main2}.
\end{align}
The derivation of Eq.~\eqref{main1} and Eq.~\eqref{main2} is presented in Section B of the supplementary material.
\begin{figure}[h]
 \centering
\includegraphics[scale=0.3]{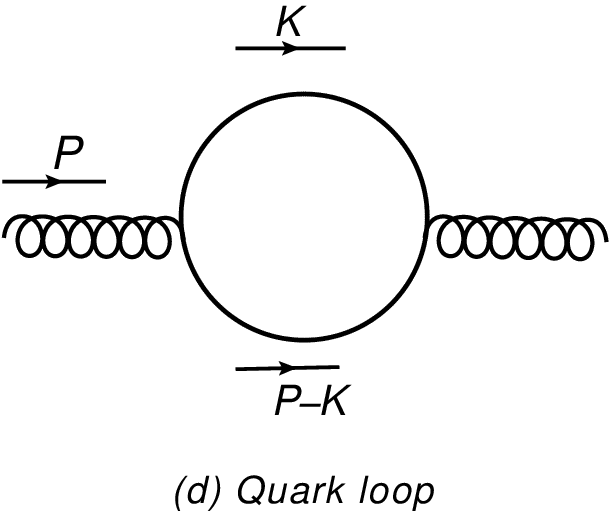}
\label{d3}
\end{figure}
\begin{figure*}
		\begin{center}
			\includegraphics[scale=0.55]{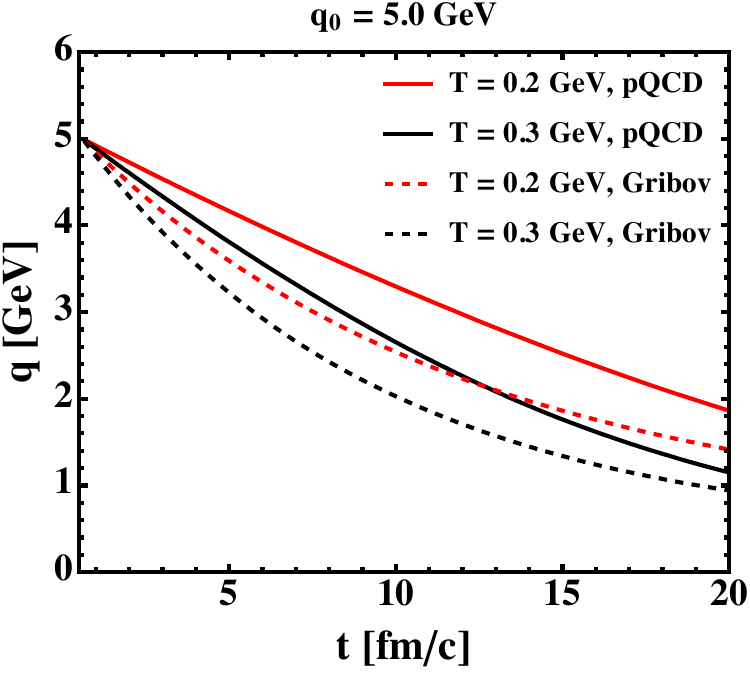}
   \hspace{1cm}
   \includegraphics[scale=0.565]{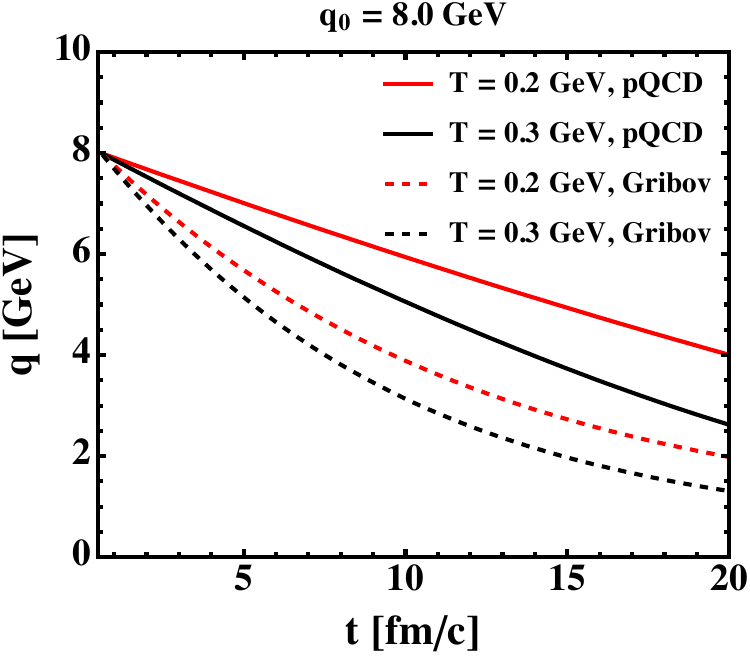}
			\caption{Momentum evolution of the massless parton in Gribov plasma with different choices of initial momentum.}
			\label{fig:drag_p}
		\end{center}
	\end{figure*}

Quark loop contribution to the gluon self-energy $\Pi_{\mu \nu}^{d}$ remains the same as it is in the usual perturbative case. Similar to the thermal case, we have
\bea
\Pi_{\mu \nu}^d(p_0,\mathbf{p})
&=& \frac{g^2 T^2}{6} N_f \int \frac{d\Omega}{4\pi} \bigg(\delta_{\mu 0} \delta_{\nu 0}+ \frac{i \omega}{P\cdot \hat{K}}\hat{K}_{\mu} \hat{K}_{\nu}\bigg),\nn
\eea
with $N_f$ as the number of flavors. Hence, we can define
\begin{align}
   &\Pi_{00}^d(p_0,\mathbf{p})=\frac{g^2T^2}{6}N_f \bigg(1- \frac{p_0}{p}R_0(p_0,p)\bigg),\\
   &\Pi_{\mu \mu}^d(P)= \frac{g^2 T^2}{6}N_f \int \frac{d\Omega}{4\pi} (1)=\frac{g^2 T^2}{6}N_f,
\end{align}
with $R_0(p_0,p)=\frac{1}{2}\rm{ln}\big|\frac{p_0+p}{p_0-p}\big|-\frac{i\pi}{2}\theta(p^2-p_0^2)$. Gribov parameter-dependent form factors can be obtained by summing up all the contributions from different sectors, and this can be used in the analysis of energy loss formalism as described in Eq.~\eqref{main}. For the quantitative estimation of the energy loss of parton in the Gribov plasma, 
we consider the temperature dependence of the $\gamma_G$ that can be described from the gap equation as~\cite{Madni:2022bea},
\bea
\gamma_G=\frac{d-1}{d} \frac{N_c}{4\sqrt{2}\pi} g^2 T.
\eea
Details of the temperature behavior of $\gamma_G$ are discussed in the supplementary material (Section C).\


{\bf \emph{Results and discussions}}- {We start with the discussion of momentum dependence of Gribov-modified parton energy loss in the QGP medium. We performed the numerical integration of the integral described in Eq.~\eqref{main} by employing the effective gluon propagator for the Gribov-Zwanziger plasma. The gluon and ghost loops contribution to the gluon self-energy is modified with the Gribov-Zwanziger framework and is reflected in the energy loss of the test parton in the medium. In Fig.~\ref{elp}, energy loss experienced by the moving massless and massive parton in the medium is plotted as a function of its momentum at a finite temperature. For the massive case, we chose M=1.25 GeV. The estimations from the Gribov-Zwanziger approach are compared with those from the perturbative calculations in the thermal medium. It is seen that non-perturbative effects incorporated through the Gribov-Zwanziger framework significantly enhance the leading order pQCD estimation of the energy loss pattern of the moving parton. The analysis holds true for both massless and massive cases. This observation is consistent with the result of the heavy quark diffusion coefficient $2\pi D T$ based on the transport theory framework in the Gribov and quasiparticle approaches~\cite{Madni:2022bea, Scardina:2017ipo} as $2\pi D T\propto \frac{1}{-dE/dx}$. Notably, the pQCD estimation of energy loss of massless and massive parton converges at a higher momentum regime. However, in Gribov plasma, where the analysis incorporates a temperature-dependent Gribov mass parameter along with the parton mass, we observe that the mass of the parton exerts a more substantial impact on the energy loss even at higher momentum scales.
}

The impact of the Gribov parameter on the momentum evolution of the massless parton is depicted in Fig.\ref{fig:drag_p}. The proper time evolution of the parton momentum is obtained by solving Langevin equations. Langevin equation ~\cite{Moore:2004tg, Rapp:2018qla,Das:2013kea} requires the knowledge of the stochastic force and dissipative force experienced by the energetic parton in the medium. The latter is estimated from the Gribov-modified energy loss, and the random kicks are described by a Gaussian noise with the width related to the momentum diffusion coefficient. We use the fluctuation-dissipation theorem to determine the momentum diffusion in the analysis~\cite{Mazumder:2013oaa,Walton:1999dy}. We observe that the parton loses a higher fraction of its 
initial momentum while propagating through the Gribov medium due to the soft interactions in comparison with that in a thermal pQCD medium. Notably, the proper time evolution of the momentum is sensitive to the value of the initial momentum of the parton and the temperature of the medium.

{We have considered the static thermal medium by fixing a constant temperature profile. This choice is justified as the focus of the current study is to analyze the non-perturbative effect of the QCD medium by comparing estimations from the Gribov-Zwanziger framework and pQCD results. Importantly, we find that the non-equilibrium corrections to parton energy loss arising due to the medium evolution are negligible when compared to the impact of non-perturbative effects. The gluon self-energy of anisotropic non-equilibrium can be decomposed as $\Pi^{ij}=\alpha A^{ij}+\beta B^{ij} + \gamma Y^{ij}+ \delta Z^{ij}$ where $Y^{ij}=\frac{\tilde a^i \tilde a^j}{\tilde a^2}$, $Z^{ij}={ p^i \tilde a^j+ p^j \tilde a^i}$ with $\tilde a^i=A^{ij}a_j$ ({\bf a} is the direction of anisotropy). The structure functions $\alpha, \beta, \gamma, \delta$ for the anisotropic QCD medium are discussed in detail in Refs.~\cite{Romatschke:2003ms,Jamal:2017dqs}. Employing the form of $\Pi^{ij}$, parton energy loss in an evolving anisotropic medium can be defined as,
 \begin{align*}
   & {\frac{d{ E}}{dx}}=-i\frac{1}{|{\bf v}|}g^2C_F v^i v^j \int\frac{d^3{\bf p}}{(2\pi)^3}\omega \bigg[\Delta_1 (\omega)(A^{ij}-Y^{ij})\nonumber\\
    &+\Delta_2 (\omega)\Big((\omega^2-|{\bf p}|^2-\alpha-\gamma)B^{ij}+(\omega^2-\beta)Y^{ij}+\delta Z^{ij}\Big)\bigg],
    \end{align*} 
with $\Delta_1=\omega^2-|{\bf p}|^2-\alpha$ and $\Delta_2=(\omega^2-\beta)(\omega^2-|{\bf p}|^2-\alpha-\gamma)-|{\bf p}|^2 \delta^2\,\tilde a^2$.
We have verified that these non-equilibrium corrections to the energy loss due to the anisotropy of the evolving medium are marginal in comparison with the non-perturbative effect of the QCD medium. This conclusion aligns with the findings of the recent study~\cite{Kurian:2020orp}, where it was shown that non-equilibrium corrections play a negligible role in the momentum evolution of charm quarks within a $1+3$-dimensional evolving QGP medium.

\begin{figure}
 \centering
 \includegraphics[scale=0.6]{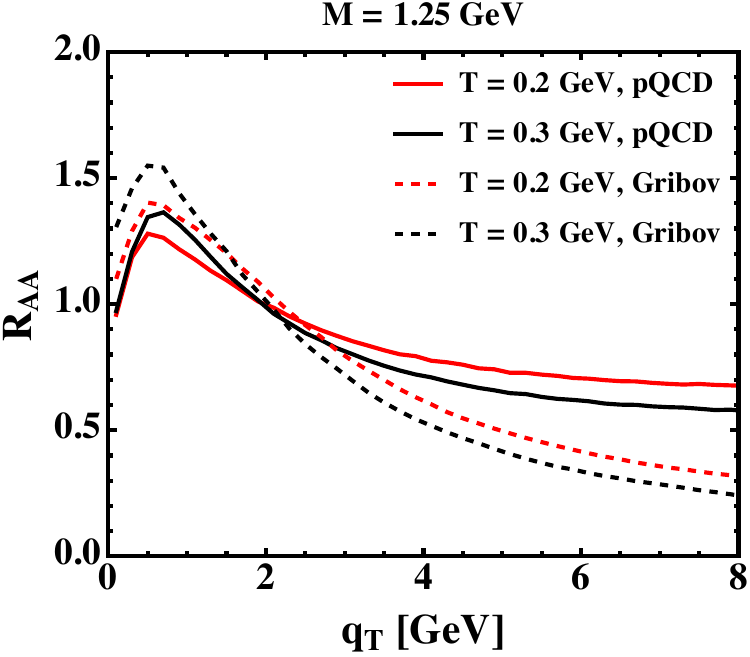}
\caption{$R_{AA}$ in the static Gribov-Zwanziger plasma as a function of transverse momentum.}
\label{RAA}
\end{figure}

{To quantify the phenomenological implication of the Gribov-modified energy loss of parton in the context of heavy-ion collision experiments, we study the nuclear suppression factor $R_{AA}(q_T)$ in a static medium that can be described as~\cite{Moore:2004tg},
\begin{align}
 R_{AA}(q_T)=\frac{f_{\tau_f} (q_T )}{f_{\tau_0} (q_T)},  
\end{align}
where $f_{\tau_0}(q_T)$ and $f_{\tau_f}(q_T)$ represent the initial and final momentum distribution of the parton in the medium at the time evolution period.  Here, we consider the case of charm quark with M=1.25 GeV and $\tau_f$=5 fm/c. The initial momentum spectra and $f_{\tau_0}$, have been taken from fixed order + next-to-leading log (FONLL) calculations, as described in Refs.~\cite{Cacciari:2005rk,Cacciari:2012ny}. Final spectra and $f_{\tau_0}$ are obtained by solving the stochastic Langevin dynamics with the Gribov-modified transport coefficients in the medium as an essential input parameter to the simulation. We observe a stronger suppression (smaller $R_{AA}$) upon the inclusion of non-perturbative effects as depicted in Fig.~\ref{RAA}, and the impact is more pronounced in the higher momentum regimes. This finding is a consequence of the fact that Gribov plasma imposes a greater hindrance for the energetic parton as it traverses through the medium. }

{\bf \emph{Summary and Outlook}}- {In this work, we studied the energy loss of a test parton, considering the non-perturbative resummation using the Gribov-Zwanziger prescription for the first time. We utilized Wong's equations along with linearized Yang-Mills equations to quantify the back-reaction exerted on the parton. This analysis involves examining the polarization effects of the medium, thereby determining the parton's energy loss as it traverses through the medium. As a first step, we estimated the gluon, quark, and ghost loop contributions to the Gribov-modified gluon self-energy. Notably, gluon and ghost loops are modified with the inclusion of the Gribov parameter, and their effects are reflected in the energy loss pattern of the energetic parton in the medium. 
We conducted a systematic analysis of the momentum evolution of the massless fast-moving parton by solving the Langevin dynamics. We observe that the Gribov-modified estimation of the energy loss pattern and momentum evolution significantly improves the leading order pQCD results. Moreover, the incorporation of non-perturbative effects is essential for ensuring consistency in the theoretical description of parton energy loss within a realistic QGP medium. Further, we explored the nuclear suppression factor for the charm quark to analyze the phenomenological implication of the present study. It is seen that $R_{AA}$ is highly sensitive to the non-perturbative effects, evidenced by a stronger suppression in the Gribov plasma in comparison with that of the leading order pQCD estimation.}

{Looking ahead, it will be interesting to extend the energy loss formalism to an evolving medium within the Gribov approach which requires the knowledge of the medium evolution of the Gribov plasma. The thermodynamics and temperature dependence of transport coefficients of the Gribov plasma have recently studied in Ref.~\cite{Jaiswal:2020qmj}. The challenging task ahead lies in obtaining second-order viscous hydrodynamics within the Gribov-Zwanziger framework, which is essential for describing the medium evolution and exploring energy loss phenomena in an expanding medium. Another interesting direction is to study the influence of fluctuations in chromodynamic fields during the early stages of evolution (which may lead to energy gain~\cite{Chakraborty:2006db, Jamal:2020fxo}) on the passage of parton within the non-perturbative resummation method. We leave these interesting aspects for future works. }\\

{\bf \emph{Acknowledgments}}- We are thankful to Najmul Haque and Santosh K. Das for the useful discussions.  M.D. is supported by the DAE, Govt. of India. 
R.G. is supported by the U.S. National Science Foundation under Grant No. PHY-2209470.
M.Y.J would like to acknowledge the SERB-NPDF (National postdoctoral fellow) File No.PDF/2022/001551. 
M.K. acknowledges the support from the Special Postdoctoral Researchers Program of RIKEN. 


\bibliography{main}

\end{document}


\section*{SUPPLEMENTAL MATERIAL}
\subsection{Tadpole diagram and gluon loop contribution $\Pi_{\mu\mu}$ and $\Pi_{00}$}\label{A0}
The Contribution of the tadpole diagram and gluon loop to the gluon self-energy can be respectively described as,
\begin{align}
	&\Pi_{\mu \nu}^a(P)= 3 C_A g^2  \int \frac{d^4K}{(2\pi)^4} \bigg(\delta_{\mu \nu} - \frac{K_{\mu}K_{\nu}}{K^2}\bigg) \Delta(K) ,\\
&\Pi_{\mu \nu}^b(P)=
	 - 6 g^2 C_A I^b_{\mu \nu}- \frac{1}{2} g^2 C_A \delta_{\mu \nu} J_1 + \frac{1}{2} g^2 C_A \delta_{\mu \nu} J_2,
\end{align}
with $C_A=N_c=3$. In Euclidean space $K^2= \omega_n^2+k^2$. The expression of $I_{\mu\nu}$, $J_1$ and $J_2$ can be defined as~\cite{Debnath:2023dhs},
\begin{align}
    I_{\mu\nu}^b(P)&= \int\frac{d^4K}{(2\pi)^4} K_\mu K_\nu\Delta(K)\Delta(P-K),\\
J_1 &= \int\frac{d^4K}{(2\pi)^4}\Delta(K) = \frac{1}{4\pi^2}\sum_{a=\pm 1} \int_0^{\infty}\frac{k^2 dk}{E_a(k)} n_B (E_a(k)),\\
J_2(P)&= \sumintb \,\frac{d^4K}{(2\pi)^4} \, i \gamma_G^2 \,\Delta(K) \Delta(P-K) \nonumber \\
&= \frac{i \gamma_G^2}{4}\sum_{a,b=\pm 1} \int\frac{d^4 K}{(2 \pi)^4} \frac{a}{[\omega_n^2 + E_a^2 (k)]} \frac{1}{[(\omega - \omega_n)^2 + E_b^2 (|\mathbf{p}-\mathbf{k}|)]} \nonumber\\
&=-\frac{i\gamma_G^2}{16}\int \frac{d^3k}{(2\pi)^3} k^2 \sum_{a,b=\pm} a \frac{n_B(E_a(k))+n_B(E_b(k))}{E_a(k) \,E_b(k)} \(-\frac{2}{ E_a(k)+E_b(k)}\) \nonumber \\
&+ \frac{i\gamma_G^2}{16} \int \frac{dk}{(2\pi)^2} k^2 \sum_{a,b=\pm1}a\frac{n_B(E_a(k))-n_B(E_b(k))}{k E_a(k)} \frac{1}{p}\[Q_0(i\omega_1,p)-Q_0(i\omega_2,p)\] \nonumber \\
&- \frac{i\gamma_G^2}{16T}\int \frac{dk}{(2\pi)^2}k^2 \sum_{a,b=\pm1}a \frac{n_B(E_b(k))(1+n_B(E_b(k)))}{E_a(k) E_b(k)} 2\[Q_1(i\omega_1,p)+Q_1(i\omega_2,p)\],
\end{align}
with $E_a(k)=\sqrt{k^2+ia\gamma_G^2}$. The net contribution from the tadpole and gluon loop for the longitudinal and transverse parts can be expressed as,
\begin{align}
\Pi_{00}^{a+b}(p_0,{\bf p})&= -\frac{13}{2}g^2 C_A J_1+ 3 g^2 C_A J_{4}(p_0, {\bf p})+ \frac{13}{2}g^2 C_A J_2(p_0, {\bf p})+6g^2 C_A J_3(p_0,{\bf p}),\\
 \Pi_{\mu \mu}^{a+b}(p_0,{\bf p})&=\frac{5}{2} g^2 C_A (\delta_{\mu \mu}-1) J_1+ \frac{1}{2}g^2 C_A \delta_{\mu \mu} J_2(p_0, {\bf p})- 6g^2 C_A I_{\mu \mu}^b(p_0, {\bf p}),
\end{align}
where $I_{\mu\mu}^b=J_1 -J_2$ and $I_{00}^b=J_1 - J_2 - J_3$. The term $J_3$ and $J_4$ can be defined as,
\bea
J_3(p_0,\mathbf{p})&=& \sumintb \,\frac{d^4K}{(2\pi)^4} k^2 \Delta(K) \Delta(P-K), \\
&=& - \frac{1}{16}\int \frac{d^3k}{(2\pi)^3} k^2 \sum_{a,b=\pm} \frac{n_B(E_a(k))+n_B(E_b(k))}{E_a(k) \,E_b(k)} \(-\frac{2}{ E_a(k)+E_b(k)}\)  \nonumber \\
&& +\frac{1}{16} \int \frac{dk}{(2\pi)^2} k^4 \sum_{a,b=\pm1}\frac{n_B(E_a(k))-n_B(E_b(k))}{k E_a(k)} \frac{1}{p}\[Q_0(i\omega_1,p)-Q_0(i\omega_2,p)\] \nn \\
&&- \frac{1}{16T}\int \frac{dk}{(2\pi)^2}k^4 \sum_{a,b=\pm1} \frac{n_B(E_b(k))(1+n_B(E_b(k)))}{E_a(k) E_b(k)} 2\[Q_1(i\omega_1,p)+Q_1(i\omega_2,p)\], \\
J_4(p_0,\mathbf{p}) &=& \sumintb \,\frac{d^4K}{(2\pi)^4} k^2 \sum_{a=\pm} \frac{1}{K^2 (K^2+ i a \gamma_G^2)}
= \frac{1}{4\pi^2}\sum_{a=\pm 1} \int_0^{\infty} dk\frac{k^4 \, n_B (E_a(k))}{E_a(k)(- a i \gamma_G^2)} .
\eea

\subsection{Ghost loop contribution of $I^c_{\mu\mu}$ and $I^c_{00}$ }\label{A}
We can express $I^c_{\mu \mu}$ as,
\begin{align}
I^c_{\mu \mu} &=\sumintb \frac{d^4 K}{(2 \pi)^4}\frac{K^2}{K^4 (P-K)^4} \nonumber\\
&= \frac{\partial}{\partial a}\bigg[\int \frac{d^3k}{(2\pi)^3} \frac{1}{4 E_1 E_2}\bigg(1-n_f(E_1)-n_f(E_2)\bigg)\bigg(\frac{1}{i\omega - E_1 -E_2}-\frac{1}{i \omega + E_1 + E_2}\bigg)\bigg]\nonumber\\
&-\frac{\partial}{\partial a}\bigg[\int \frac{d^3k}{(2\pi)^3} \frac{1}{4 E_1 E_2}\bigg(n_f(E_1)-n_f(E_2)\bigg)\bigg(\frac{1}{i\omega + E_1- E_2}-\frac{1}{i\omega -E_1+ E_2}\bigg)\bigg],
\label{App:imumu}
\end{align}
with $E_1=k$, $E_2=\sqrt{(p-k)^2+a}\approx \sqrt{k^2 +a}- \frac{\mathbf{p}\cdot \mathbf{\hat{k}}}{\sqrt{k^2+a}}$. Here, $n_f(E_1)=\frac{1}{e^{k/T}+1}$ and employing the definition of $E_2$, we can express $n_f(E_2)=n_f(\sqrt{k^2+a}) - \mathbf{p}\cdot \mathbf{\hat{k}} \frac{\partial}{\partial k}n_f(\sqrt{k^2+a})$. We have,
\be
\label{A1}
   I^c_{\mu \mu}= \mathcal{J}_1 + \mathcal{J}_2.
\ee
Employing the definition of $E_1$, $E_2$, $n_f(E_1)$, and $n_f(E_2)$ we have,
\begin{align}
\label{A2}
    \mathcal{J}_1 &= \frac{\partial}{\partial a}\Big[\int \frac{d^3k}{(2\pi)^3} \frac{1}{4 E_1 E_2}(-n_f(E_1)-n_f(E_2))\Big(\frac{1}{i\omega - E_1 -E_2}-\frac{1}{i \omega + E_1 + E_2}\Big)\Big] \nonumber \\
    &= \frac{\partial}{\partial a}\Big[\int dk\, \frac{4\pi k^2}{(2\pi)^3} \frac{1}{4 k \sqrt{k^2+a}}\frac{2(n_f(k)+n_f(\sqrt{k^2+a}))}{k+\sqrt{k^2+a}}\Big]\nonumber\\
   &=  - \frac{1}{4\pi^2} \int dk\,  \frac{e^{k/T}k+3T (1+  e^{k/T})}{4(1+e^{k/T})^3 k^3 T} \nonumber\\
   &= \frac{1}{4\pi^2}\int dk\, \Big[\frac{1}{4k^2}\frac{\partial n_f(k)}{\partial k}- \frac{3 n_f(k)}{4k^3}\Big]\nonumber \\
   &= - \frac{1}{16 \pi^2} \frac{7 \zeta(3)}{8 \pi^2 T^2}, 
\end{align}
\begin{align}
    \mathcal{J}_2 &= -\frac{\partial}{\partial a}\Big[\int \frac{d^3k}{(2\pi)^3} \frac{1}{4 E_1 E_2}(n_f(E_1)-n_f(E_2))\Big(\frac{1}{i\omega + E_1- E_2}-\frac{1}{i\omega -E_1+ E_2}\Big)\Big] \nonumber\\
   &=- \frac{\partial}{\partial a} \int dk\, \frac{2 \pi k^2}{(2\pi)^3} d(\cos \theta ) \,  \frac{1}{4 k \sqrt{k^2 +a}}\Big(n_f(k)- n_f(\sqrt{k^2+a})+ p \cos \theta \frac{\partial n_f(\sqrt{k^2+a})}{\partial k}\Big) \nonumber\\
   &\times \Big(\frac{1}{i \omega + k - \sqrt{k^2+a}+ \frac{p k \cos \theta}{\sqrt{k^2+a}}}-\frac{1}{i \omega - k + \sqrt{k^2+a}- \frac{p k \cos \theta}{\sqrt{k^2+a}}} \Big) \nonumber \\
   &=- \frac{\partial}{\partial a}  \int \frac{dk\, d\cos \theta \, k^2}{(2\pi)^2} \frac{1}{4k\sqrt{k^2+a}} (n_f(k)- n_f(\sqrt{k^2+a})) \nonumber \\
   &\times \frac{\sqrt{k^2+a}}{k} \Big(\frac{1}{\frac{\sqrt{k^2+a}}{k}(i \omega + k- \sqrt{k^2+a})+ p \cos \theta } -\frac{1}{\frac{\sqrt{k^2+a}}{k}(i\omega - k+ \sqrt{k^2+a})-p \cos \theta}\Big)  \nonumber \\
   &- \frac{\partial}{\partial a} \int \frac{dk\, d\cos \theta \, k^2}{(2\pi)^2} \frac{1}{4k\sqrt{k^2+a}} p \cos \theta \frac{\partial n_f(\sqrt{k^2+a})}{\partial a} \frac{\sqrt{k^2+a}}{k}\nonumber \\
   &\times \Big(\frac{1}{\frac{\sqrt{k^2+a}}{k}(i \omega + k- \sqrt{k^2+a})+ p \cos \theta } -\frac{1}{\frac{\sqrt{k^2+a}}{k}(i\omega - k+ \sqrt{k^2+a})-p \cos \theta}\Big)  \nonumber\\
   &= - \frac{\partial }{\partial a} \Big[ \int \frac{dk\, k^2}{(2\pi)^2} d \cos \theta \, \frac{n_f(k)- n_f(\sqrt{k^2+a})}{4 (k^2)} \Big(\frac{1}{i\omega^1 + p \cos \theta}-\frac{1}{i\omega^2 - p \cos \theta}\Big)\Big] \nonumber \\
   & - \frac{\partial }{\partial a}\Big[ \int \frac{dk \, d \cos \theta }{(2\pi)^2} \frac{k^2}{4(k^2)} \,  p \cos \theta \frac{\partial n_f(\sqrt{k^2+a})}{\partial k}\Big(\frac{1}{i\omega^1 + p \cos \theta}-\frac{1}{i\omega^2 - p \cos \theta}\Big)\Big]\nonumber\\ 
   &\approx - \frac{\partial}{\partial a} \Big[\int \frac{d\cos\theta \, dk}{(2\pi)^2} \frac{e^{k/T}a}{8k^3 T (1+e^{k/T})^2}\Big(\frac{1}{p_0+ p \cos \theta}- \frac{1}{p_0- p \cos \theta}\Big)\Big] \nonumber\\
   &- \frac{\partial}{\partial a} \int \frac{dk\, d\cos\theta\,  k^2}{(2\pi)^2} \frac{1}{4 k^2} p \cos \theta \frac{\partial}{\partial k}\Big(n_f(k)-\frac{a}{2k T}n_f(k)(1-n_f(k))\Big)\nonumber\\
   &\times \Big(\frac{1}{p_0+ p \cos \theta}
    - \frac{1}{p_0- p \cos \theta} + a \Big[\frac{k-p_0}{2k^2 (p_0+ p \cos \theta)^2}+ \frac{k+p_0}{2k^2 (p_0- p \cos \theta)^2}\Big]\Big) \nonumber \\
 &\approx 0- \frac{\partial}{\partial a}\Big[\int \frac{dk}{(2\pi)^2} k^2 \frac{a}{4}\Big(- \frac{1}{k^2}\frac{\partial}{\partial k}\Big(\frac{n_f(k)(1-n_f(k))}{2 k T}\Big)\Big) d(\cos \theta) \Big(\frac{1}{p_0 + p \cos \theta}- \frac{1}{p_0 - p \cos \theta}\Big) \Big]\nonumber\\
 & -\frac{\partial}{\partial a}\Big( \int \frac{dk \, k^2}{(2\pi)^2} p \cos \theta \, d(\cos \theta) \, \frac{a}{4 k^2} \frac{\partial n_f(k)}{\partial k}\Big[\frac{k-p_0}{2k^2(p_0 + p \cos \theta)^2}+ \frac{k+p_0}{2k^2(p_0 - p \cos \theta)^2}\Big]\Big).
 \end{align}
 Hence, we obtain
\begin{align}
\label{A3}
 \mathcal{J}_2= - \frac{1}{4\pi^2} \frac{7 \zeta(3)}{8 \pi^2 T^2}\Big(\frac{p_0^2}{p_0^2-p^2}- \frac{p_0}{2p} \log\Big(\frac{p_0+p}{p_0-p}\Big)\Big).
\end{align}
 Utilizing Eq.~\eqref{A2} and Eq.~\eqref{A3} in Eq.~\eqref{A1}, we obtain
 \be
 I^c_{\mu \mu}= - \frac{1}{16\pi^2} \frac{7 \zeta (3)}{8 \pi^2 T^2}- \frac{1}{4\pi^2} \frac{7 \zeta(3)}{8 \pi^2 T^2}\left(\frac{p_0^2}{p_0^2-p^2}-\frac{p_0}{2p}\log \left(\frac{p_0+p}{p_0-p}\right)\right).
 \ee
Now, we compute the ghost loop contribution of $I_{00}$. We have,
\bea
I^c_{00}&=&\int \frac{d^4K}{(2\pi)^4} \frac{k_0^2}{(K^2+a)^2 ((P-K)^2+b)^2} = \lim_{a,b \to 0} \frac{\partial^2}{\partial a \partial b} \Big(\frac{k_0^2}{(k_0^2+k^2+a)((p_0-k_0)^2+(p-k)^2+b)}\Big) \nonumber \\
&=& \lim_{a,b \to 0} \frac{\partial^2}{\partial a \partial b} \Big(\frac{1}{(P-K)^2+b}- \frac{k^2+a}{(k_0^2+k^2+a)((p_0-k_0)^2+(p-k)^2+b)}\Big) \nonumber \\
&=& \lim_{a,b \to 0} \frac{\partial }{\partial a} \frac{\partial }{\partial b} \int \frac{d^3K}{(2\pi)^3}   \frac{k^2+a}{E_a E_b}  (1- n_f(E_a)-n_f(E_b))\Big(\frac{1}{i\omega -E_a - E_b}- \frac{1}{i\omega + E_a + E_b}\Big) \nonumber \\
&&- (n_f(E_a)-n_f(E_b))\Big(\frac{1}{i\omega + E_a- E_b}-\frac{1}{i\omega -E_a + E_b}\Big) \nonumber\\
&=& \mathcal{T}_1 + \mathcal{T}_2.
\eea
Employing the forms $E_a=\sqrt{k^2+a}$, $E_b= \sqrt{(p-k)^2+b}\approx \sqrt{k^2+b}- \frac{p k \cos \theta}{\sqrt{k^2+b}}$, $\mathcal{T}_1$ and $\mathcal{T}_2$ can be further simplified as,
\bea
\mathcal{T}_1 &=& \lim_{a,b \to 0} \frac{\partial ^2}{\partial a \partial b} \left[\int \frac{d^3k}{(2\pi)^3} \frac{k^2+a}{E_a E_b} \frac{2 (n_f(E_a)+ n_f(E_b))}{E_a+ E_b}\right]\\ 
&=& \lim_{a,b \to 0} \frac{\partial ^2}{\partial a \partial b} \int \frac{4\pi k^2 \, dk}{(2\pi)^3} \frac{\sqrt{k^2+a}}{\sqrt{k^2+b}} 2 \frac{n_f(\sqrt{k^2+a})+n_f(\sqrt{k^2+b})}{\sqrt{k^2+a}+\sqrt{k^2+b}} \nonumber \\
&=& \int \frac{dk}{\pi^2}k^2 \frac{e^{k/T}-T- T e^{k/T}}{8(1+e^{k/T})^2 k^5 T} = \int_0^{\infty} \frac{dk}{\pi^2} \frac{1}{8k^3} \left[k \frac{\partial n_f(k)}{\partial k}- n_f(k)\right] 
= - \frac{1}{8 \pi^2} \frac{7 \zeta(3)}{8 \pi^2 T^2},
\eea
\bea
\mathcal{T}_2 &=& - \lim_{a,b \to 0}\frac{\partial^2}{\partial a \partial b} \int \frac{d^3k}{(2\pi)^3} \frac{k^2+a}{\sqrt{k^2+a}\sqrt{k^2+b}} \Big(n_f(\sqrt{k^2+a})-n_f(\sqrt{k^2+b})+ p \cos \theta \frac{\partial n_f(\sqrt{k^2+b})}{\partial k}\Big) \nonumber \\
&\times& \Big(\frac{1}{i\omega + \sqrt{k^2+a}-\sqrt{k^2+b}+ \frac{p \, k \cos \theta}{\sqrt{k^2+b}}}- \frac{1}{i\omega - \sqrt{k^2+a}+\sqrt{k^2+b}- \frac{pk \cos \theta}{\sqrt{k^2+b}}}\Big) \nonumber \\
&=& \mathcal{T}_{21} + \mathcal{T}_{22},
\eea
where
\bea
\mathcal{T}_{21}&=& -\lim_{a,b \to 0} \frac{\partial^2}{\partial a \partial b}\int \frac{d^3k}{(2\pi)^3} \frac{k^2+a}{\sqrt{k^2+a}\sqrt{k^2+b}} \left(n_f(\sqrt{k^2+a})-n_f(\sqrt{k^2+b})\right) \nonumber \\
&\times& \left(\frac{1}{i\omega + \sqrt{k^2+a}-\sqrt{k^2+b}+ \frac{p k \cos \theta}{\sqrt{k^2+b}}}- \frac{1}{i\omega - \sqrt{k^2+a}+\sqrt{k^2+b}- \frac{p k \cos \theta}{\sqrt{k^2+b}}}\right) \nonumber \\
&=& \frac{1}{(2\pi)^2}\frac{1}{p_0^2-p^2},
\eea
and
\bea
\mathcal{T}_{22} &=& -\lim_{a,b \to 0} \frac{\partial^2}{\partial a \partial b}\int \frac{d^3k}{(2\pi)^3} \frac{k^2+a}{\sqrt{k^2+a}\sqrt{k^2+b}}  p \cos \theta \frac{\partial n_f(\sqrt{k^2+b})}{\partial k}\nonumber \\
&\times& \left(\frac{1}{i\omega + \sqrt{k^2+a}-\sqrt{k^2+b}+ \frac{pk \cos \theta}{\sqrt{k^2+b}}}- \frac{1}{i\omega - \sqrt{k^2+a}+\sqrt{k^2+b}- \frac{pk \cos \theta}{\sqrt{k^2+b}}}\right) \nonumber \\
&\approx & \frac{1}{(2\pi)^2}\frac{p^2}{(p^2- p_0^2)^2}- \frac{2}{(2\pi)^2} \frac{7 \zeta(3)}{8\pi^2 T^2} \left(\frac{p_0^2}{p_0^2-p^2}- \frac{p_0}{2p}\log \left(\frac{p_0+p}{p_0-p}\right)\right). 
\eea

\subsection{Extracting the Gribov parameter }
 The Gribov-modified parton energy loss in the medium has a dependence on the temperature behaviour of the Gribov parameter. The Gribov approach modifies the partion function and thereby improves the Faddeev-Popov quantization procedure as,
\bea
\mathcal{Z}&=& \int \mathcal{D}A\,\mathcal{D}\bar c\,\mathcal{D}c\, \text{det}[-\partial\cdot A]\, \delta(\partial\cdot A) \,V(\Omega)\nonumber\\
&&\hspace{-.5cm} \times \exp\bigg[-S_{\text{YM}}-\int d^4x\, \bar c^a(x)\partial_\mu D_\mu^{ab}c^b(x)\bigg],
\label{YM}
\eea
where $c$ and $\bar c$ represent the ghost and anti-ghost fields, respectively. Here, $D_\mu^{ab}$ is the covariant derivative and $S_{\text{YM}}$ denotes the Yang-Mills action. The  no pole condition which prohibits the path integral over field configurations from extending beyond the Gribov horizon $\Omega=\{A: \partial \cdot A=0, -\partial \cdot D\ge 0\}$ is enforced with the inclusion of the step function $V(\Omega)$ with the form as follows~\cite{Vandersickel:2012tz, Fukushima:2013xsa},
\bea
V(\Omega)&=&\theta[1-\sigma(P=0)]
=\int_{-i\infty +\epsilon}^{+i\infty +\epsilon} \frac{e^{s(1-\sigma[0])}}{2\pi i s}ds.
\eea
Introducing the mass parameter $\gamma_G$ that minimizes the exponent of the integral Eq.~\eqref{YM} at specific value of $s$, one can obtain gap equation at finite temperature for Gribov mass parameter as~\cite{Vandersickel:2012tz}, 
\begin{align}
g^2\frac{(d-1)N_c}{d}\sumintb_P{\frac{1}{P^4+\gamma_G^4}}=1,
\end{align}
with $d$ as the space-time dimension and Euclidean four-momentum $P$ temporal component exhibits discrete bosonic Matsubara frequencies. In the  asymptotic limit ($T\rightarrow \infty$), temperature dependence of the $\gamma_G$ can be described from the gap equation as~\cite{Madni:2022bea},
\bea
\gamma_G=\frac{d-1}{d} \frac{N_c}{4\sqrt{2}\pi} g^2 T.
\eea
It is important to emphasize that the temperature behaviour of the Gribov parameter is critically depending on the running coupling constant. The running coupling is taken as,
\bea
g^2(T)=\frac{48\pi^2}{11 N_c-2N_f}\frac{1}{\ln(\Lambda^2/\Lambda_{\overline{\text{MS}}}^2)},
\eea
where $\Lambda_{\overline{\text{MS}}}$ is extracted from the lattice estimation $\alpha_s(\Lambda=1.5\, \text{GeV}, N_f=3)=0.326$~\cite{Bazavov:2012ka}. We consider $T_c=160$ MeV for the quantitative estimations in the analysis.

\bibliography{main}